\documentstyle[12pt,psfig]{article}
\begin{document}

\hfill{SLAC-PUB-7973}

\hfill{October 1998}

\begin{center}
{\Large {\bf A coupled channel unitary chiral approach to the
meson-meson interaction and $\pi \pi$ scattering in a nuclear 
medium$^1$}}
\end{center}

\vspace{0.2cm}

\begin{center}
{\large {E. Oset $^2$, J.A. Oller$^2$, J.R. Pel\'aez$^3$, Chiang Huan
Ching$^4$ and M.J. Vicente Vacas$^2$
  }}
\end{center}

\vspace{0.3cm}

{\small {\it
$^2$ Departamento de F\'{\i}sica Te\'orica,  Universidad
de Valencia and  IFIC, Centro Mixto Universidad
de Valencia - CSIC, 46100 Burjassot (Valencia), Spain.

$^3$ Stanford Linear Accelerator Center. Stanford University, Stanford,
California 94309.

$^4$ Institute of High Energy Physics, Chinese Academy of Sciences,
Beijing 100039 China.
}}

\vspace{.3cm}

\begin{abstract}
A new tool which combines chiral perturbation theory and unitarity in
coupled channels is applied with success to the study of the meson-meson
interaction, extending the theoretical predictions up to
$\sqrt{s} = 1.2$ GeV, hence improving considerably the convergence
radius  of conventional chiral perturbation theory, $\chi PT$. The method
is applied to obtain the meson-meson phase shifts and inelasticities 
as well as the isoscalar $\pi \pi$ scattering amplitude inside a nuclear
medium.
\end{abstract}
\footnotetext[1]{Research partially supported
by the Department of Energy under contract DE-AC03-76SF00515
and Spanish CYCIT AEN-0776 and PB96-0753 }

\vspace{.3cm}

\begin{center}
Invited talk presented at

 {\bf Conference 
on Mesons and Light Nuclei '98} 

Prouhonic, Prague, Czechoslovakia,
August 31-September 4, 1998.
\end{center}

\vspace{.5cm}

The starting point of the approach is unitarity in coupled channels, which in
matrix form is most easily stated in terms of the real $K$ matrix
as 

\begin{equation}
T^{-1} = K^{-1} \, - i\, \sigma,
\end{equation}
where $T$ is the scattering matrix and $\sigma$ is a diagonal
matrix that measures the phase space available for the intermediate
states

\begin{equation}
\sigma_{nn} (s) = - \frac{k_n}{8 \pi \sqrt{s}}\, \theta \, \left(s -
(m_{1n} + m_{2n})^2\right),
\end{equation}
where $k_n$ is the on shell CM momentum of the meson in the
intermediate state $n$ and $m_{1n}, m_{2n}$ the masses of the two
mesons in the state $n$.
The  meson-meson states considered here are $K \bar{K},
\pi  \pi, \eta \eta, \pi \eta, \pi K , \pi \bar{K},  \eta K$ and $
\eta\bar{K}$.

From eq. (1) one immediately realizes that 

\begin{equation}
K^{-1} = \hbox{Re} \, T^{-1},
\end{equation}
and hence we have in matrix form 

\begin{equation}
T = [\hbox{Re}\, T^{-1} \, - i\, \sigma]^{-1},
\end{equation}

The next step is to make an expansion of $\hbox{Re}\,T^{-1}$ in
powers of $p^2$, like in $\chi P T$. Remarkably, the expansion 
of $\hbox{Re}\,T^{-1}$ has better chances of
convergence, since $T$ has poles (the "$\sigma$"  in $I = 0$
appears around 500 MeV)  and perturbation theory will
necessarily break before that. Similarly one could not
get the other meson-meson resonances like the $f_0 (980)$
for $J = 0, I = 0$, or the $\rho$ and $K^*$ vector mesons, etc....
In contrast, where $T$ has poles  $T^{-1}$ will have zeros, 
which, in principle, do not give any convergence problem, thus
allowing us to obtain resonances.

The expansion of $\hbox{Re}\, T^{-1}$ is also suggested by the
analogy with the effective range formula in Quantum
Mechanics, which states that (in an s-wave elastic channel, for simplicity)

\begin{equation}
K^{-1} = \sigma \, \hbox{ctg} \, \delta  \, \propto - \frac{1}{a} + \frac{1}{2}
\, r_o k^2,
\end{equation}
with $k$ the one particle momentum, $a$ the scattering length and $r_0$
the effective range.

In $\chi P T$, $T^{-1}\simeq T_2^{-1}(1-T_4\,T_2^{-1}...)$,
where $T_2$ is the lowest order, $O (p^2)$, amplitude and $T_4$ is just the
$O(p^4)$ term. In order to avoid problems with the inversion of $T_2$,
 we multiply eq. (4) by $T_2 T_2^{-1}$ on the right 
and $T_2^{-1} T_2$ on the left. That is

\begin{equation}
T = T_2\, [T_2\, \hbox{Re}\, T^{-1}\, T_2 - i\, T_2\, \sigma\, T_2]^{-1}\, T_2.
\end{equation}
In addition, an 
ordinary expansion of $T_2 \,\hbox{Re}\, T^{-1}\, T_2$ in powers of
$p^2$ gives

\begin{equation}
T_2 \,\hbox{Re}\, T^{-1}\, T_2 \simeq T_2 - \hbox{Re}\, T_4 \cdots
\end{equation}
Thus, recalling that in the physical  region $\hbox{Im}\, T_4 = T_2 \,\sigma
\,T_2$, our scheme provides, up to $O (p^4)$

\begin{equation}
T = T_2\, [T_2 - T_4]^{-1}\, T_2,
\end{equation}
which is a generalization of the inverse amplitude method
of \cite{1} to coupled channels. 
The standard $ O (p^2)$ and $O (p^4)$ Lagrangians  of
Gasser and Leutwyler \cite{2} are used to evaluate $T_2$ and
$T_4$. Those calculations are lengthy, but even if they are not complete, 
we can still use some approximations. We have several ways to proceed:

\begin{itemize}
\item[a)] If a full calculation of $T_4$ with 
one loop in the $s$, $t$ and $u$
channels and tadpoles is avaliable, a straightforward
application of eq. (8) is possible  \cite{3}. The
method allows for a direct comparison of the
$L_i$ coefficients of the $O (p^4)$ Lagrangians, which are
fitted to the meson-meson data, with those of the
standard $\chi P T$ expansion.

\item[b)] A simpler, equally successful scheme, is obtained by omitting
crossed loops and tadpoles. Their effect is
 reabsorbed in the $L_i$ coefficients,
which are also fitted to the data.
In such case we can use the loop function (symmetric matrix) in
the s-channel containing two meson propagators

\begin{equation}
G_{nn} (s) = i \, \int \frac{d^4 q}{(2 \pi)^4} \frac{1}{q^2
- m_{1n} + i \epsilon} \, \frac{1}{(P - q)^2 - m_{2n}^2 + i \epsilon},
\end{equation}
with $P$ the total momentum of the two meson system. It satisfies

\begin{equation}
\hbox{Im} \, G_{nn} (s) = \sigma_{nn}.
\end{equation}

The real part of $G$ is obtained by means of a suitable cut off in
$|\vec{q}|$ that makes the loop convergent. Dimensional regularization is
equally suitable. Changes of the cut off revert in changes in the $L_i$ 
coefficients and the final solution for the meson-meson amplitudes
is cut off independent. Thus, in this approach, we take

\begin{equation}
\hbox{Re}\, T_4 = T_2  \, \hbox{Re} \, G \, T_2 +  T_4^p,
\end{equation}
where $T_4^p$ is a polynomial obtained from the tree level 
contribution of the $O (p^4)$ chiral Lagrangians.

This is the method followed in \cite{4}.

\item[c)] Finally a third option comes from assuming that for some
particular cut off $T_2 \, \hbox{Re}\, G \, T_2$ in eq.  (11) can make
the $T^p_4$ contribution negligible. This is not possible 
for all the meson channels but it was shown to work in the scalar
sector ($J= 0)$ in \cite{5}. In this latter case we would obtain

\begin{equation}
T = T_2\, [T_2 - T_2\, G\, T_2]^{-1} \, T_2 = [1 - T_2\, G]^{-1}\, T_2,
\end{equation}
or, equivalently,

\begin{equation}
T = T_2 + T_2 \, G \, T,
\end{equation}
which is nothing but the Bethe-Salpeter equation, with $T_2$ playing the role
of a potential, and where $T_2$ and $T$ are factorized outside the
$d^4 q$ integral in the $T_2 \, G \, T$ term of eq. (13), a 
feature that was shown in \cite{5} using different arguments.
The cut off needed for a good description of the scalar data is
around 1 GeV.

\end{itemize}

In fig. 1 we show some results which are obtained with the option
b) discussed above. We fitted the values of the $L_i$ to the
data, but $L_6$ and $L_8$ only appear in the combination
$2 L_6 + L_8$. Hence one has 7 parameters in the theory.
The agreement with the data is quite good, and all the resonances
(their masses, widths and associated poles) are well reproduced.

\begin{figure}
\hbox{
\psfig{file=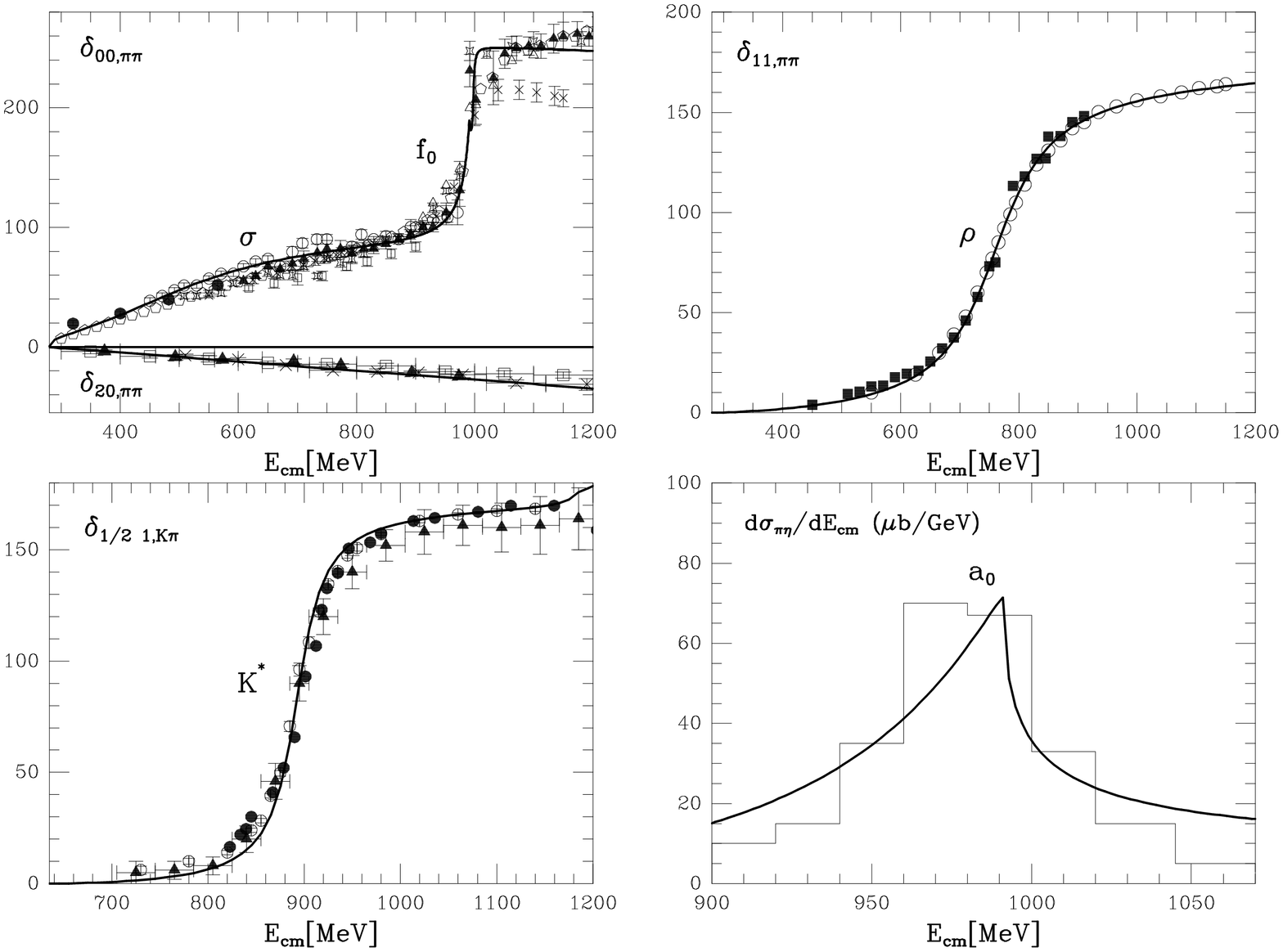,width=15cm,angle=-90}}
\hbox{
\psfig{file=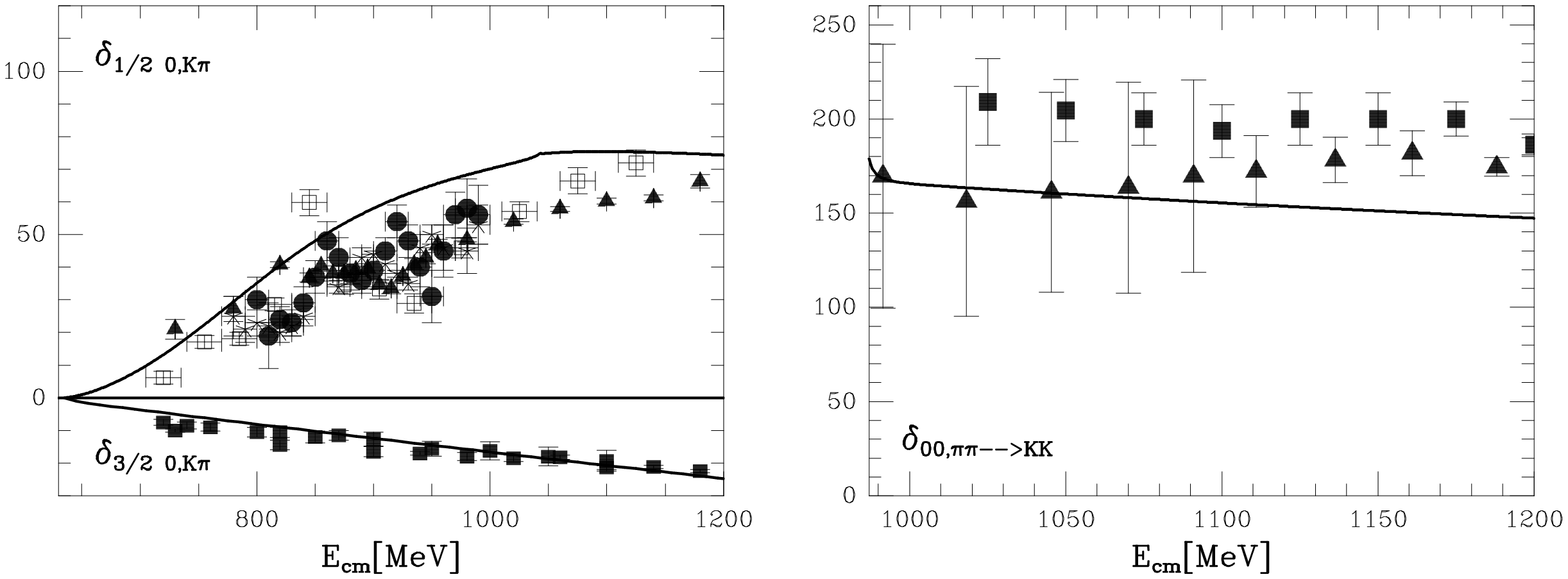,width=15cm,angle=-90}
}
{\bf Figure 1:}{We display the results of method b)
for the phase shifts of $\pi\pi$
scattering in the $(I,J)=(0,0),(1,1),(2,0)$ channels, where the $\sigma$, 
$f_0$ and $\rho$ resonances appear, together with those of
$\pi\pi\rightarrow K \bar K$, as well as the phase shifts of $\pi K$
scattering in the $(3/2,0),(1/2,0) $ and $(1/2,1)$ channels, where 
we can see th
appearance of the $K^*$ resonance. The results also include
the $\pi^-\eta$ mass distribution 
for the $a_0$ resonance in the $(I,J)=(1,0)$ channel from
$K^-p\rightarrow \Sigma(1385)\pi^-\eta$.
For reference to the data, see \cite{4} and references therein.}
\end{figure}

Next we discuss a subject of much interest in the interface between
meson-meson interactions and nuclear physics, which is the $\pi \pi$ 
interaction in a nuclear medium.

The topic was initiated in \cite{6}, where, due to the interaction
of the pions with the nucleus, the $J=0$ $\pi \pi$ interaction 
developed some strength below threshold, leading to pairs
of bound mesons, a kind of pion Cooper pairs. Further studies
imposing minimal chiral constraints in the amplitudes softened that
strength and the singularities below threshold do not appear
\cite{7}. Yet, this accumulation of strength close to threshold,
whereas the vacuum amplitude vanishes, 
could have some observable
consequences. Indeed, in \cite{8} the invariant mass distribution for
$\pi^+ \pi^-$ close to threshold was appreciably enhanced with
respect to that of $\pi^+ \pi^+$ in the study of ($\pi, 2 \pi)$
reactions in nuclei.

We have performed a recent calculation using the previous method
for the $\pi \pi$ interaction \cite{9}. It is
interesting to see that the off shell dependence of the $\pi \pi$
amplitude cancels out with some other diagrams and, hence, only
the on shell $\pi \pi$ amplitudes in vacuum 
are needed as input \cite{10}.

In fig. 2 we show the results for $\hbox{Im} \, T_{00}$ in $J = I = 0$
in $\pi  \pi$ scattering for different values
of the Fermi momentum, $k_F$. We can appreciate that as $k_F$  
increases some strength accumulates at low energies  around and
below threshold which could explain the experimental increase \cite{8}
in the invariant mass distribution.

\begin{figure}
\hbox{
\psfig{file=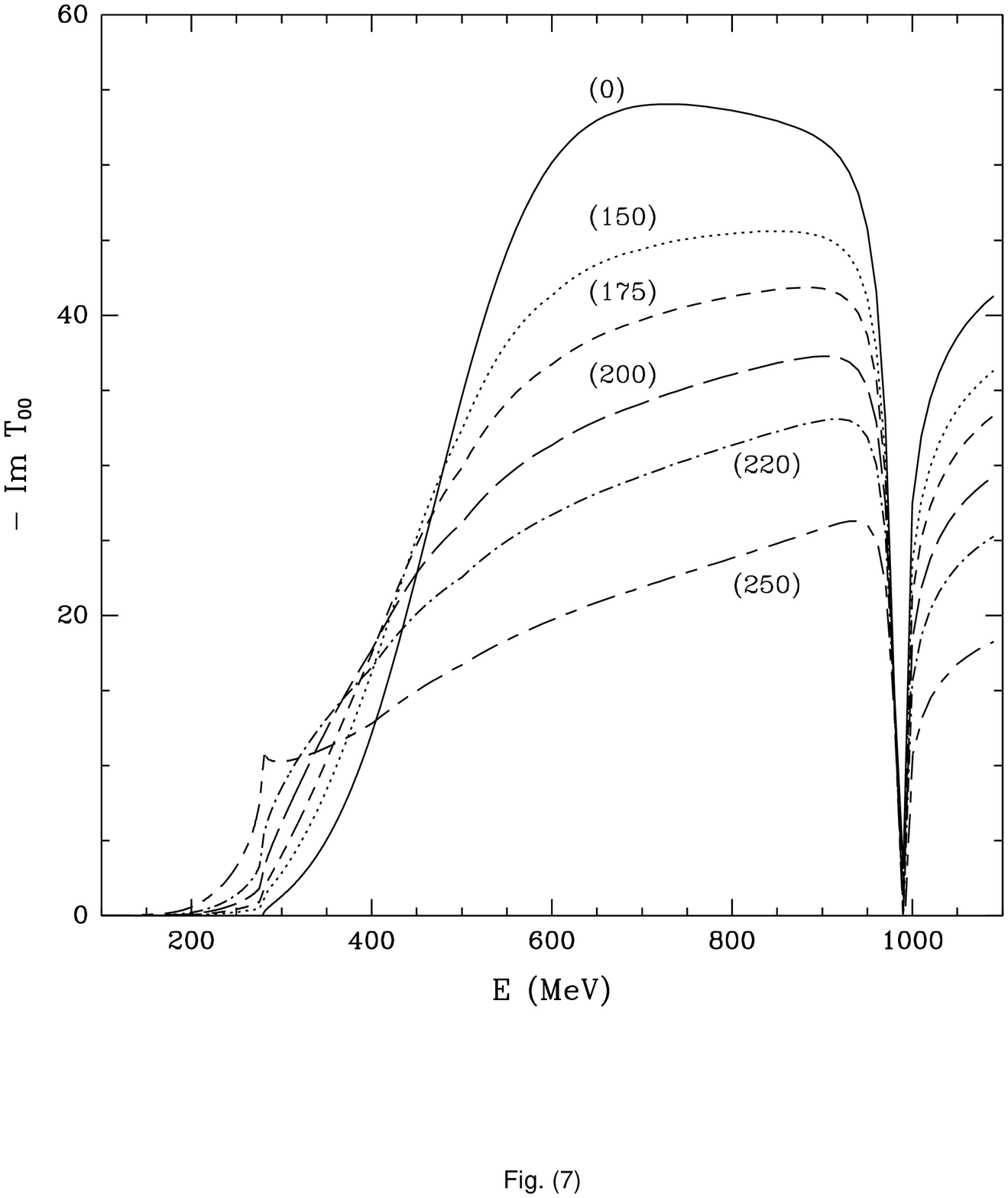,width=0.8\textwidth}}

{\bf Figure 2:} {Im$T_{22}$ for $\pi\pi\rightarrow \pi\pi$ scattering in
$J=I=0$ ($T_{00}$ in the figure) in the nuclear medium for different values of
$k_F$ versus the CM energy of the pion pair. The labels correspond to the
values of $k_F$ in MeV.}
\end{figure}

Summarizing, we have found a technique similar to that
of the effective range in Quantum Mechanics, which 
improves the convergence of standard $\chi P T$. The approach 
yields all the meson resonances below 1.2 GeV, 
which cannot be obtained with standard $\chi P T$. 
The application of the method to the study
of the scalar isoscalar $\pi \pi$ interaction in a nuclear medium leads
to an enhanced strength around the two pion threshold which
could explain present invariant $\pi^+ \pi^-$ mass distribution
measured in the $(\pi, 2 \pi)$ reaction in nuclei.

\end{document}